\documentclass[12pt]{article}
\input{psfig}
\usepackage{graphics}
\usepackage{ifthen}

\newcommand\jpbgraphics[2]{
			   \newpage 
			   \ifthenelse{ \equal{#2}{} }
                               {
  			       \resizebox{\textwidth}{!}
                                         {\includegraphics{#1}}
                               }
                               { \begin{figure}
                                     \resizebox{\textwidth}{!}
                                         {\includegraphics{#1}}
                                     \caption{#2}
                                 \end{figure}
                               }
                           }

\newcommand{\beq}{\begin{equation}}
\newcommand{\eeq}{\end{equation}}
\newcommand{\beqa}{\begin{eqnarray}}
\newcommand{\eeqa}{\end{eqnarray}}

\newcommand{\nm}{\nonumber\\}

\begin{document}

\title{How fast does Langton's ant move?} 
\author{Jean Pierre Boon \thanks{E-mail address: {\tt jpboon@ulb.ac.be}}\\
{\em Center for Nonlinear Phenomena and Complex Systems} \\
{\em Universit\'{e} Libre de Bruxelles, 1050 - Bruxelles, Belgium}}
\date{\small(submitted to the Journal of Statistical Physics - April 2000)}

\maketitle

\begin{abstract}
The automaton known as `Langton's ant' exhibits a dynamical transition from 
a disordered phase to an ordered phase where the particle dynamics (the
ant) produces a regular periodic pattern (called `highway'). Despite the
simplicity of its basic algorithm, Langton's ant has remained a puzzle in
terms of analytical description. Here I show that the highway dynamics 
obeys a discrete equation where from the speed of the ant ($c={\sqrt 2}/52$) 
follows exactly. 

\vspace*{.1in}

\noindent
{\bf Keywords}: Langton's ant; propagation; lattice models.

\end{abstract}

Langton's ant has been a recurring theme in the mathematical and physical
literature \cite{gale}. There are two reasons. The first is of physical
relevance: the automaton known as Langton's ant (which I describe below)
offers a prototype of complexity out of simplicity, and, in particular, 
can be viewed as a model of dynamical transition from a disordered phase 
to an ordered phase~\cite{langton}. The second reason is mathematical: 
despite the simplicity of the basic algorithm, the spatio-temporal dynamics 
generated by the automaton (see Fig.1) has so far resisted analytical 
treatment.

The basic process governing the automaton dynamics follows a simple rule.
The automaton universe is the square lattice with checker board parity
so defining H sites and V sites. A particle moves from site to site (by
one lattice unit length) in the direction given by an indicator. One may
think of the indicator as a `spin' (up or down) defining the state of the 
site. When the particle arrives at a site with spin up (down), it is 
scattered to the right (left) making an angle of $+\pi/2$ ($-\pi/2$) with 
respect to its incoming velocity vector. But the particle modifies the
state of the visited site (up $\Longleftrightarrow$ down) so that on its
next visit, the particle is deflected in the direction opposite to the 
scattering direction of its former visit. Thus the particle entering from 
below a H site with spin up is scattered East, and on its next visit to 
that same site (now with spin down), if it arrives from above, it will be 
scattered East again, while if it arrives from below, it will be scattered
West. Similar reasoning shows how the particle is scattered North or South
on V sites.

At the initial time, all sites are in the same state (all spins up or all 
spins down), and the position and velocity direction of the particle are 
fixed, but arbitrary. So if we paint the sites black or white according to 
their spin state, we start initially with say an all white universe. Then 
as the particle moves, the visited sites turn alternately black and white 
depending on whether they are visited an odd or even number of times. This 
color coding offers a way to observe the evolution of the automaton universe. 
The particle starts exploring the universe by first creating centrally 
symmetric transient patterns (see figures in references \cite{gale}), 
then after about 10 000 time steps (9977 to be precise), 
it leaves a seemingly `random territory'\footnote{The disordered 
phase is not what a random walk would produce: the automaton is deterministic 
and its rules create correlations between successive states of the substrate, 
so also between successive positions of the particle. The power spectrum 
computed from the particle position time correlation function measured over 
the first 9977 time steps goes like $\sim \nu^{-\zeta}$ with 
$\zeta \simeq 4/3$. In the ordered phase (`highway'), the power spectrum 
shows a peak at $\nu = 1/104$ with harmonics.} to enter a `highway' 
(see Fig.1) showing a periodic pattern: in the highway, the particle 
travels with constant propagation speed.

A theorem by Bunimovich and Troubetzkoy \cite{bunimov} demonstrates that 
the automaton fulfills the conditions for unboundedness of the trajectory 
of the particle (see the highway in Fig.1). But, to the best of our 
knowledge, the dynamical transition from the disordered phase to the ordered 
phase remains unexplained to date.\footnote{As well as the transient 
symmetrical patterns in the early stages of the dynamics.} 
Here, I consider the ordered phase and show analytically that the 
propagation speed is $c = \sqrt 2/52$ (in lattice units) as measured in 
automaton simulations \cite{gale}.   

Because of the complexity of the dynamics on the square lattice, Grosfils,
Boon, Cohen, and Bunimovich \cite{grosfils} developed a one-dimensional
version of the automaton for which they provided a complete mathematical
analysis also applicable to the two-dimensional triangular lattice.
One of their main results is the mean-field equation describing the
microscopic dynamics of the particle subject to the more general condition
that the spins at the initial time are randomly distributed on the lattice.
The equation reads, for the one-dimensional lattice, 
\beq
\label{1d_eq}
f(r+1,t) \,=\, q\, f(r,t-1)\,+\,(1-q)\, f(r,t-3)\,,
\eeq
and, for the two-dimensional triangular lattice,
\beqa
\label{2d_eq}
f(r+1,t)&=& q(1-q)\, f(r,t-2) \nm
        &+& [q^2 +(1-q)^2]\, f(r,t-8)  \nm
        &+& (1-q)q\, f(r,t-14)\,,
\eeqa
where $f(r,t)$ is the single particle distribution function, i.e. the
probability that the particle visits site $r$ {\em for the first time}
at time $t$, and $q$ is the probability that the immediately previously
visited site along the propagation strip (the highway) has initially 
spin up, i.e. the probability that the particle be scattered, in the 
one-dimensional case, along the direction of its velocity vector when 
arriving at the scattering site at $r-1$, and, in the two-dimensional 
triangular case, along the direction forming clockwise an angle of 
$+2\pi/3$ with respect to the incoming velocity vector of the particle. 
Equations (\ref{1d_eq}) and (\ref{2d_eq}) express the probability of a
first visit at a site along the propagation strip in terms of the
probability of an earlier visit at the previous site along the strip.
Note that in the two-dimensional case, the equation describes the
one-dimensional propagation motion along the edge of the strip. 
The equations were shown to yield exact solutions for propagative 
behavior (corresponding to an ordered phase of the lattice) in the two 
classes of models considered by Grosfils {\em et al.} \cite{grosfils}.

Equations (\ref{1d_eq}) and (\ref{2d_eq}) are generalized as follows
\beqa
\label{gen_eq}
f(r+\rho,t)\, =\, \sum_{j=0}^{n}\, p_j(q)\; f(r,t-\tau_j) \;\;; \nm
\tau_j\, =\, (1 + \alpha j)\, m\, \tau\;\;;\;\;  \sum_{j=0}^{n} p_j =1\;\;;
\;\; n \leq r/\rho \;,
\eeqa 
where, as above, $f(r,t)$ is the {\em first visit} distribution function.
Here $\rho$ denotes the elementary space increment of the dynamics along 
the propagation strip. $p_j$ is the probability that the particle propagates
from $r$ to $r + \rho$ in $\tau_j$ time steps, i.e. $\tau_j$ is the time 
delay between two successive first visits on the strip (more precisely on 
the one-dimensional edge of the strip) for the path with probability~$p_j$, 
and $m$ is the corresponding minimum number of automaton time steps 
($\tau_0~=~m\tau$, where $\tau$ is the automaton time step, $\tau =1$). 
The sum is over all possible time delays, weighted by the probability~$p_j$ 
(a polynomial function of $q$). 
$\alpha$ denotes the number of lattice unit lengths in an `elementary loop', 
i.e. the minimum number of lattice unit lengths necessary to return to 
a site.\footnote{An interesting equation follows from the continuous limit of
(\ref{gen_eq}); this is discussed elsewhere (J.P. Boon, to be published).}

Now from the expectation value of the time delay, computed with (\ref{gen_eq}),
\beq
\label{exp_tau}
E[\tau (q)]\,=\,\sum_{j=0}^{n}\,\tau_j\, p_j(q)\,=\,(1 + \alpha\,
\langle j \rangle )\,m\,\tau\,,
\eeq 
where $\langle j \rangle = \sum_{j=0}^{n}\,j\, p_j(q)$, one obtains 
immediately  the average propagation speed of the particle in the ordered 
phase:  $c(q) = \rho / E[\tau (q)]$. 
 
It is straightforward to verify that Eqs. (\ref{1d_eq}) and (\ref{2d_eq})
are particular cases of the general equation (\ref{gen_eq}): \\
for 1-D : $\alpha = 2,\, m = 1,\, \rho =1,\, n = 1\,,$ with $p_0 = q  \,,\,
 p_1 = 1 - q\,$;\\
for 2-D (triangular lattice) :   
$\alpha = 3,\, m = 2,\, \rho =1,\, n = 2\,,$ 
with $p_0 = q\,(1 - q) \,,\, \\
p_1 = q^2 + (1 - q)^2 \,,\, p_2  = (1 - q)\,q\,$. \\
The corresponding propagation speeds are then readily obtained from 
(\ref{exp_tau}); for the one-dimensional case one finds
$c(q) = 1/\langle \tau(q) \rangle = [1 + 2(1-q)]^{-1}= 1/(3-2q)$, and
for the triangular lattice: $\langle \tau(q) \rangle =
[1+3(q^2 + (1-q)^2 + 2q(1-q))]\times 2 $, so that $c=1/8$. These results
are in exact agreement with those obtained in~\cite{grosfils}.

For the 2-D square lattice : $\alpha = 4,\, m = 2 \times 4,\, 
\rho = 2 \sqrt 2\,$. The value of $\rho$ is easily checked by inspection 
of the highway path shown in the upper box of Fig.1 : it is the length of 
the elementary increment along the edge of the propagation strip. 
Correspondingly $m$ is $2 \times 4\,$ (because the minimum number of time 
steps necessary to move one elementary space increment must be counted on 
each edge of the strip). 
For the square lattice, one does not know the value of $n\,,$ but from the 
structure of the $p_j$'s for the 1-D and 2-D triangular lattices given 
above, one can infer a plausible value: $n = 6$, with 
$p_0 = p_6 = q^2\,(1-q)^2 \,,\,p_1 = p_5 = q\,(1-q)\,[q^2\,+\,(1-q)^2]\,,\,
p_2 = p_4 = p_0 + p_1\,,\,p_3 = [q^2\,+\,(1-q)^2]^2\,$. However the precise 
expressions are unimportant for the automaton describing Langton's ant, 
because all sites are initially in the same spin state; so $q = 1$, 
and only one $p_j$ is non-zero: $p_3 = 1$. 
Equation~(\ref{gen_eq}) then reads:
\beqa
\label{2d_sq_eq}
f(r+2 \sqrt 2\,,t)\, =\, f(r,t-\tau_3) \;\;; \nm 
\tau_3\, =\, (1 + 4 \times 3)\,2 \times 4\,=\,104\,,
\eeqa
which describes the dynamics of the particle in the highway. This result
shows that a displacement of length $2 \sqrt 2$ along the edge of the
strip is performed in $104$ automaton time steps. Consequently the
propagation speed of Langton's ant in the highway is 
$c = \rho / \tau_3 = 2 \sqrt 2 /104 = \sqrt 2 /52$. 

The origin of particle propagation in 1-D and 2-D triangular lattices 
was shown to be a `blocking mechanism' \cite{grosfils}, and the 
question was raised as to whether such a mechanism also exits in the square 
lattice. Although the precise blocking mechanism has yet to be identified, 
that the same general equation, Eq.(\ref{gen_eq}), describes propagation in 
1-D, 2-D triangular and square lattices suggests that a similar blocking 
mechanism is responsible for propagative dynamics in the highway of 
Langton's ant.

\bigskip

{\em Note}: In reference \cite{grosfils}, the `reorganization corollary'
for the 2-D triangular lattice (Corollary 3, p.599) was incorrectly stated. 
It should read: {\em All sites located on one edge of the propagation strip 
are in the initial state of the sites on the other edge, shifted upstream by 
one lattice unit length}. The particle dynamics can then be interpreted as
the controller of a Turing machine which transcribes and shifts the string of 
characters (0's and 1's for L and R) of the input tape (on one edge) to the 
output tape (on the other edge). The control operator is the EXCHANGE
gate of Feynman's model of a quantum computer \cite{feynman}. 
The same corollary applies trivially to the spin states (up and down spins 
interchanged as 0's and 1's) on the edges of the highway of Langton's ant.

\bigskip

{\large \bf{Acknowledgments}}.

\bigskip

I enjoyed discussions with P. Grosfils, E.G.D. Cohen, D. Meyer and \
O.~Tribel, and I acknowledge support by the {\em Fonds National de la 
Recherche Scientifique} (FNRS, Belgium).

\newpage
\thispagestyle{empty}

\jpbgraphics{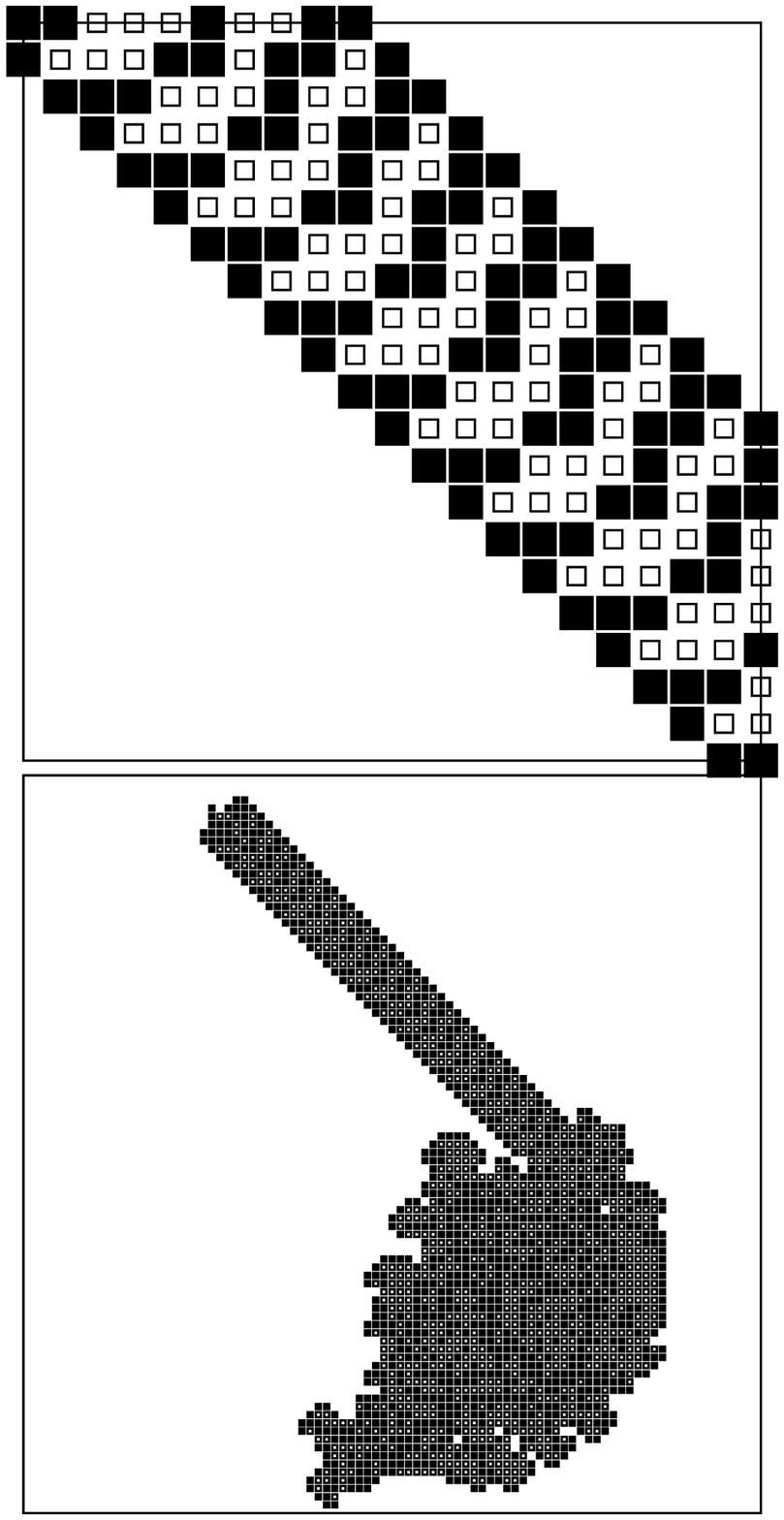}  
{Langton's ant trajectory after 12,000 automaton time steps. The upper
box is a blow-up of the highway showing the periodic pattern. Sites with
open squares and dark squares have opposite spin states (up and down).}

\end{document}